Gerontologic Biostatistics 2.0: Developments over 10+ years in the age of data science

Chixiang Chen PhD;[1,2] Michelle Shardell, PhD;[1,3] Jaime Lynn Speiser, PhD, MS;[4] Karen Bandeen-Roche, PhD;[5] Heather Allore, PhD;[6] Thomas G Travison, PhD;[7,8] Michael Griswold, PhD;[9] Terrence E. Murphy, PhD[10]

[1]Department of Epidemiology and Public Health, University of Maryland School of Medicine; [2]Department of Neurosurgery, University of Maryland School of Medicine; [3]Institute for Genome Sciences, University of Maryland School of Medicine; [4]Department of Biostatistics and Data Science, Wake Forest University School of Medicine; [5]Departments of Biostatistics, Medicine and Nursing, Johns Hopkins University; [6]Department of Internal Medicine, Yale School of Medicine and Department of Biostatistics Yale School of Public Health; [7]Marcus Institute for Aging Research, Hebrew Senior Life; [8]Division of Gerontology, Beth Israel Deaconess Medical Center, Harvard Medical School; [9]Departments of Medicine and Data Science, University of Mississippi Medical Center; [10]Deparment of Public Health Sciences, Penn State College of Medicine

**Corresponding Author:**
Michelle Shardell, PhD
Institute for Genome Sciences
Department of Epidemiology and Public Health
670 W Baltimore Street
Baltimore, Maryland 21201

(T) 410-706-1136

Email: mshardell@som.umaryland.edu

**ABSTRACT**

**Background**: Introduced in 2010, the sub-discipline of gerontologic biostatistics (GBS) was conceptualized to address the specific challenges in analyzing data from research studies involving older adults. However, the evolving technological landscape has catalyzed data science and statistical advancements since the original GBS publication, greatly expanding the scope of gerontologic research. There is a need to describe how these advancements enhance the analysis of multi-modal data and complex phenotypes that are hallmarks of gerontologic research.

**Methods**: This paper introduces GBS 2.0, an updated and expanded set of analytical methods reflective of the practice of gerontologic biostatistics in contemporary and future research.

**Results:** GBS 2.0 topics and relevant software resources include cutting-edge methods in experimental design; analytical techniques that include adaptations of machine learning, quantifying deep phenotypic measurements, high-dimensional -omics analysis; the integration of information from multiple studies, and strategies to foster reproducibility, replicability, and open science.

**Discussion**: The methodological topics presented here seek to update and expand GBS. By facilitating the synthesis of biostatistics and data science in gerontology, we aim to foster the next generation of gerontologic researchers.

**INTRODUCTION**

The past decade has witnessed explosive growth in the depth and breadth of gerontological research in areas such as geroscience, multimorbidity, sarcopenia, aging in place, dementia, long-term care, and more. The health implications of these topics, advancements in data-generating technology, and increased interest in in precision medicine in aging have motivated corresponding developments in biostatistics (1, 2). As a result, there is a substantial demand for the analysis of large-scale data, a pressing need for analytical approaches (3) that provide a more nuanced understanding of health and aging, and a need for integration of biostatistical efforts with those of the diverse set of fields collectively known as 'data science' in gerontology research (4).

The sub-discipline of gerontologic biostatistics (GBS) was introduced in 2010 to emphasize the special challenges in the analysis of research studies of older adults (5); since the original publication, multiple statistical advancements in this age of data science have facilitated gerontologic research of greater breadth and depth. This paper aims to update and enrich GBS by presenting recent analytic advancements for data-rich gerontologic research. These advancements include innovations in experimental design and feature analytic approaches such as machine learning, deep phenotypic measurement and high-dimensional -omics analysis that collectively enhance gerontology, geroscience, and geriatric research. Other advancements include the growing integration of information from multiple studies and several promising means of promoting reproducibility, replicability, and open science.

Our overall goals are to enhance and expand the scope and practice of GBS to meet the demands of the big data era and to inform biostatisticians and other researchers dedicated to the study of older adults of these innovations. Moreover, we aim to foster the training of the next generation of researchers by providing an updated description of the current state of gerontologic biostatistics and data science in hopes of facilitating development of new methods in this area.

## UP-TO-DATE CHALLENGES AND INNOVATIONS IN GBS 2.0

### Advances in Experimental Design

In this era of big data, experimental design continues to play a central role in conducting causal inference in gerontologic research. While the designs used in gerontological research overlap with those of observational and interventional studies, they also possess distinctive features (5). Designs for gerontologic research need to acknowledge that older adults are at higher risk of mortality and attrition, sources of bias that cannot be prevented by randomization (6). Although the randomized controlled trial (RCT) is considered the gold standard for testing therapeutic efficacy, older adults were historically excluded from therapeutic trials. This exclusion, which has since been prohibited by the 21st Century Cares Act (7), was not limited to age and often applied to common aging-related factors such as comorbid illness, concurrent medication use, and functional disability. This discarded historical practice illustrates how screening criteria intended to improve precision and prevent missing data may inadvertently lead to biases and lack of generalizability. Happily, there are effective designs that specifically address these concerns in trials that evaluate palliative and end-of-life care, as well as deprescribing in later life (8). For improved transparency, the SPIRIT (Standard Protocol Items: Recommendations for Interventional Trials) (9) and CONSORT (10) guidelines for clinical trial protocols were written to improve the design and reporting of trials. There have also been numerous large multi-site therapeutic RCTs designed to emphasize the needs of older participants (11, 12). Moreover, a policy statement from the Food and Drug Administration emphasized their interest in encouraging a broad population sample in the development of new drugs (13). National and international oncology societies have also called for greater inclusion of older adults in trials (14-16). The National Institute on Aging in conjunction with the Alzheimer's Association has provided guidance for Alzheimer's Disease trial designs (17).

Over the past decade there have been many advances that make trials shorter, less resource intensive, and increasingly adaptive regarding allocation of participants to non-efficacious interventions. Some of these even allow for the addition of new interventions during the trial (18). These newer trial designs generally fall under adaptive designs (19), platform trials (3, 20), or dynamic regimes (21, 22). Dynamic regime trials (23) and "standardly-tailored multicomponent" trials often differ from factorial or fractional factorial designs in that participants need not be eligible for all components of the intervention to qualify for inclusion. Methods developed to tackle these complex study designs involve novel structural (causal) modeling via machine learning (21-23). Dynamic regime trials are often well suited to studies of precision medicine that seek to emphasize each patient's needs and responses to previous treatments (24). These designs also attempt to address the heterogeneity found among older adults, another important factor on the path toward evaluation of the personalization of results.

One persistent challenge is that many non-pharmacologic interventions cannot be concealed from the participant (i.e. exercise), implying that only the outcome assessor and analyst can be blinded. Another limitation is that reports rarely describe whether the components of multicomponent interventions may have directly modified the underlying risk factor; it is therefore often unknown whether the intervention worked via its direct effect on specific risk factors. Standardly-tailored multicomponent designs have been used with individual randomization trials (25), clustered randomization trials (26), and in pilot studies of older adults (27). Unfortunately, interventions found to be efficacious in RCTs often fail to exhibit effectiveness when applied in real-world settings. This translation gap from trial to practice is often attributed to the greater heterogeneity and constraints of real-world healthcare systems. To bridge this gap, pragmatic trials and implementation science designs have been developed. The National Institutes of Health's Healthcare Systems Collaboratory provides guidance to address some of the analytic and design challenges of embedding interventions in healthcare systems (28, 29). This has led to the formation of thematically focused Collaboratories, such as the National Institute on Aging Imbedded Pragmatic Alzheimer's Disease and Alzheimer's Disease-Related Dementias Clinical Trials (IMPACT) Collaboratory, which supports the embedding of non-pharmacological trials in healthcare systems (30). Despite the conduct of pragmatic trials, many of which randomize the intervention at the cluster level, there remain unanswered questions as to whether existing analytic strategies for highly controlled RCTs suffice to address the increased heterogeneity, attrition, and decreased adherence characteristic of gerontologic research. A systematic review of available software for designs such as multi-arm multi-stage trials and platform trials is presented in Meyer et al (31) (see eTable 1).

**Analytical Developments**

In addition to developments in experimental design, there have been multiple impactful technological innovations that made their mark on the practice of gerontologic biostatistics over the last 10 years. These include advancements and adaptations of machine learning (ML), quantification of phenotypic measurements, and the proliferation of –omics data and analysis methods in aging research.

**1. Gerontologic Biostatistics Welcomes Machine Learning**

In the past decade, machine learning (ML) (32) has made its way into GBS and continues to serve as a complement, supplement, and sometimes an alternative to more conventional forms of statistical modeling. ML is a powerful analytic approach and can be thought of as the capacity of a computer to process and evaluate data beyond simple instructions or programmed algorithms. For more than twenty years a great deal has been written comparing conventional statistical modeling and ML (33).

In a posting from 2018 titled 'Road Map for Choosing between Statistical Modeling and Machine Learning,' Harrell provides a high-level comparison of these disciplines and how the research questions can help guide methodologists to choose the most appropriate techniques (34). Figure 2 summarize some of Harrell's points while deliberately presenting the techniques within each discipline along a continuum of (arguably) increasing conceptual and computational complexity. In general, conventional statistical modeling may be more appropriate when the goal is to perform inference and quantify uncertainty of associations, when an additive model is a reasonable assumption, when intuition and interpretability are prioritized, and when sample sizes are modest or small. In contrast, ML, which is more closely related to computer and data science, may be preferred when discovery is at an early stage, when uncertainty of associations is not a concern, when highly complex non-linear models with higher order interactions among predictors are a necessity, when prediction or classification of outcomes is more important than inference of associations between individual covariates and outcomes, and when samples of individuals are large. Table 1 lists several high-level considerations when comparing statistical modeling and ML and some examples from the literature. In the supplement, eTable 1 presents some of the most popular R packages for ML as well as the approach offered by SAS.

The trade-offs between statistical modeling and ML in gerontology are illustrated in two recent papers on serious fall-related injury. Speiser et al (35) chose two ML methods that are among the most interpretable, i.e., decision trees and random forests, using data from the Lifestyle Interventions and Independence for Elders (LIFE) study (36). Decision trees provide an intuitive graphical algorithm that a healthcare provider can easily follow to quantify an individual's injury risk, whereas random forests quantify the relative magnitude of each variable's importance in predicting fall-related injury. Because both ML techniques provide reasonable performance with clinically intuitive results, this work provides an accessible introduction to the use of ML in GBS. In contrast, Womack et al (37) describe a hybrid approach where ML is used for classification and conventional statistical modeling is used for prediction. Specifically, the authors build a prediction model for serious falls in the Veterans Affairs Birth Cohort using multivariable logistic regression, among the most common forms of statistical modeling. The outcome of medically serious falls was classified using a previously validated support vector machine (38), a ML method that enhanced the detection of fall-related injury. This application of a well validated ML approach identified twice as many serious fall outcomes than the traditional sources without a commensurate rise in false negatives, thereby boosting predictive performance. This second study combined enhanced detection of a hard-to-detect outcome by ML with the probabilistic transparency of logistic regression. The studies outlined here demonstrate that ML and conventional statistical models can be reasonably employed to build risk prediction models of important gerontologic outcomes. With automated data collection creating increasingly complex types and quantities of data, ML is used in a growing array of gerontologic studies. Examples include novel causal inference applications, such as the construction of inverse probability weights in marginal structural models (39) and the accommodation of heterogeneous treatment effects by causal forests (40). In any case, research with either statistical modeling, ML, or both is strengthened by content-expert knowledge that facilitates discovery of reproducible and replicable findings needed to improve outcomes in older adults.

## 2. Phenotypic Measurement Challenges and Approaches

Phenotypic measurement is a fundamental component of gerontologic research that involves the quantification and characterization of observable traits and clinical features in older adults. In modern aging studies, researchers frequently grapple with a spectrum of challenges during the collection and interpretation of phenotypic data involving temporal changes and multifaceted traits. In this section, we explore the challenges and methods associated with two types of quantitative information that have become prevalent in modern gerontology research practices involving older adults; these are signal intensive measurements and multi-dimensional geriatric outcomes.

Signal intensive measurements have become common in both research and clinical settings. Often these measurements document behaviors—e.g., physical activity via accelerometry (41), sleep via polysomnography (42), and mood via ecological momentary assessment (43). These measurements may also document cellular or physiological health—e.g. metabolomic profiling. Statistical engagement has contributed greatly to signal detection and summarization of such data from older adults, and conversely data on aging have spurred important statistical developments. Real-time monitoring is amenable to functional data analysis, both non-parametric and parametric. Extensive details were provided in an outstanding review by Di et al appeared in 2019 (44). Analyses employing cosine function bases have been widely employed when diurnal variation is of interest (45). Non-parametric functional regression and principal components analysis have followed the paradigm elucidated by Ramsay and Silverman (46). The next frontier for signal-intensive data on aging is the development of methods for the joint analyses of multi-modal signals (47). These involve approaches such as the joint and individual variance explained method (48), a principal-components-based extension that decomposes the original multi-modal data into sets of components capturing (i) joint variation among the modes, (ii) variation specific to each mode, and (iii) residual noise (see Table 1 and eTable 1).

Geriatric outcomes are frequently multidimensional and may map to constructs, i.e., health states, that are theoretically rather than physically defined, and may consequently lack a single accurate/precise form of measurement (49). Such outcomes, including frailty, physical disability, and cognitive functioning, must be assessed through a multivariate measure. Frailty is a signature example. The term is recognizable--any of us can readily envisage a "frail" older person. Frailty is important for its promise to identify those persons most vulnerable to adverse outcomes, so that these outcomes can be prevented or ameliorated (50). Rigorous statistical reasoning is crucial to the assessment of frailty because operationalizing the definition has proven challenging, and different assessment methods identify "frail" persons with considerable discordance (51). One major schema (of two) has followed an "index" rubric, characterizing frailty as the proportion among a large number of "deficits"—diseases, disability in various tasks, symptoms, and family history risks experienced by an older adult (52). The other is grounded in a "scale" rubric, beginning with a definition not unlike the consensus version given above, and then proposing "criteria" by which to infer the frailty status indirectly as a physical phenotype (53). In either case, validation is then crucial to provide evidence that the measure targets what it intends. Bandeen-Roche and colleagues have recently explicated this process in the frailty setting (50). Latent variable modeling is a statistical technique that can be particularly helpful to this end. Its utility applies in good measure to the evaluation of how well one's measures comply with a construct, to the harmonization of disparate measures, and to forms of identifying and addressing biases in measurement (54).

## 3. Analyzing High-Dimensional -Omics Data with Focus on Microbiome

Advances in high-throughput technologies have contributed to an explosion in the availability of molecular –omics data in aging research, including in large epidemiologic studies of older adults. These data include transcriptomics, metabolomics, proteomics, the microbiome, and more; and research with multiple types of -omics data are beginning to show promise for identifying mechanistic pathways (55), metabolic alterations (56), and biomarker signatures (57) of aging-related phenotypes.

While these data typically share multiple statistical challenges stemming from batch effects, limits of detection, excess zeroes, and high dimensionality; we focus discussion on microbiome data due to its additional analytical challenges and hypothesized role in the biology of aging (58).

Microbiome data are both particularly challenging and an area of growing interest in gerontology, as demonstrated by the 2020 Special Issue in the *Journal* on Gut Microbiome and Aging (59-64) and other recent *Journal* articles published since (65-69). Microbial profiles are determined by sequencing the 16S ribosomal RNA (rRNA) gene and classifying the taxon of each rRNA 'read.' Therefore, data for each sample is a vector of taxa (e.g., species) read counts. A key salient feature of these data is *compositionality* (70), meaning that since total read counts differ by sample and are constrained by the DNA sequencer, rather than reflecting a sample's absolute microbial abundance, the taxon read measures its relative abundance. Therefore, counts are often normalized by a sample's total read count to produce taxa relative abundances that are constrained to sum to 1. Compositionality is also relevant in metagenomic sequencing, which sequences the full microbial genome and quantifies the relative abundances of microbial genes.

Findings from either sequencing technology can be biased if statistical methods do not account for compositionality of the data (70). For this reason, directly applying methods developed for other -omics count data (e.g., transcriptomics) to microbiome data is not recommended. Rather, compositional data analysis methods are needed. Early compositional data analysis techniques were motivated by low-dimensional contexts (71) and have been extended to address high-dimensional microbiome data. These methods include regularized or constrained regression analyses that employ log-ratio transformed relative abundances as explanatory variables (72), log-ratio differential abundance analysis of microbiome outcomes between groups (73), and compositionally aware measures of correlation (74) and distance (75). Recent methods have accommodated high-dimensional compositional microbiome data in formal causal inference, most notably in mediation analysis to identify microbial taxa on indirect pathways between lifestyle factors and health outcomes (76). These modern compositional data analysis methods, which can be implemented using R statistical software (see eTable 1) and require reporting in accordance with published guidelines (77), enable biostatisticians to adapt their workflows from other –omics data to the microbiome.

**Recent Advances in Information Integration**

Next, we shift the focus from describing design and analytical approaches within a single study to a discussion of data/information sharing by discussing research that leverages and combines information from multiple studies, a hallmark of the era of GBS 2.0. There has been an ongoing movement towards integrating data from multiple studies to provide analytic results that enhance those of individual studies. Integrative data analysis (IDA) (78) enables the incorporation of big data sources, such as electronic health records, Medicare claims, and national surveys. In gerontology, IDA leverages and combines data, leading to a more profound understanding of aging-related events and diseases. This involves integrating data collected across different time points, cohorts, or outcomes, thereby enabling researchers to overcome limitations, such as small sample sizes, rare diseases/exposures, or the limited generalizability of findings from individual studies (79-81).

Most IDA approaches assume the availability of raw data from multiple studies, which is an ideal scenario but may often be infeasible (82). Privacy concerns and complexities surrounding individual data sharing often necessitate accessing summary data from the literature or sharing partial information from distributed networks. However, these alternatives require the development of new IDA frameworks. Generalized meta-based techniques have been developed for integrating summary data (83, 84), while distributed learning algorithms have been proposed for sharing partial information that concurrently preserve privacy (85, 86) and promote efficient communication (87, 88). Several R packages support their use (see eTable 1). Nonetheless, there remains considerable room for developing new methods, such as the management of heterogeneity across multiple studies (89), the use of multiple outcomes (90), and the application of these tools to address real-world problems in gerontologic research.

**Reproducibility, Replicability, and Open Science**

Lastly, we address the challenges of reproducibility *within* a study and replicability *across* multiple studies. As researchers, it is exciting when new results are discovered; however, progress truly happens when results are replicated via independent studies using similar methodology. While reproducibility can be fostered by sharing data and code; there has also been a recent focus on rigor, replicability, and transparency in scientific and medical research that is relevant to gerontology (91). Rigor is typically defined as using appropriate design and analysis to conduct a study, as well as the comprehensive reporting of its results. Replicability refers to the ability to obtain consistent results using the same methods and analysis using independent data. Transparency involves the unfettered sharing of methods, design, code, and data that promotes reproducibility and can enable replicability. Together, rigor, replicability and transparency form the foundation of open science, which aims to make scientific research widely available (92). Moreover, in order to be rigorous, replicable, and transparent, open science depends on the field of biostatistics for optimal study designs, analytic methods and reporting of results, and the complementary field of data science for appropriate data and code sharing.

As the scientific community began to focus on replication, reproducibility, and open science, so did the community of gerontologic researchers. In their 2015 editorial, Editor-in-Chief of *The Gerontologist*, Rachel Pruchno and colleagues posed the question "Is gerontology in crisis?". They emphasized that statistical rigor and well-powered studies are crucial for gerontologic research (93). The authors advocated for manuscript submissions focused on replication studies, meta-analyses and literature reviews that synthesize related results across many studies. Specific methodologic challenges of rigor, replicability, and reproducibility in gerontology have been discussed in recent publications (1, 2, 94). One challenge is that gerontologic studies are often longitudinal with data collected over years or decades, which may make data sharing difficult. Additional challenges of replicating study results may be the dependence of those results on a certain population or generation such as World War II Veterans or baby-boomers, respectively.

Funding agencies and journals have recently adopted policies that facilitate open science, especially regarding data sharing. For example, in January 2023, the National Institutes of Health issued a Data Management and Sharing policy that requires all investigators to prospectively plan for the management and sharing of scientific data that includes the following details: what scientific data will be required for sharing, when data need to be shared (at minimum, at the time of publication of results), and standards for data quality. Many resources provide recommendations for open data sharing, including data storage, integration, and governance (95). Repositories and other resources that support open science are provided in eTable 1. In summation, there are many resources available for data and code sharing to improve the rigor, replicability, and transparency of gerontologic research, as well as enable the growing trend of collaborative and open science; how well one uses these resources may soon determine the influence of their research.

**DISCUSSION**

GBS 2.0 comprises cutting-edge statistical and data science approaches designed to meet the demands of gerontology research in the big data era. Delineating some notable advances since the original 2010 GBS publication can serve as an educational resource for biostatisticians and other researchers in gerontology. In summary, the inherent complexity of population heterogeneity requires more advanced research designs; machine learning becomes invaluable in cases where conventional statistical models prove inadequate; sophisticated methods are essential for tackling the challenges associated with temporal and multifaced phenotypic measurements; -omics data necessitate efficient methods that accommodate their challenging features; and data/information integration requires interdisciplinary solutions at the nexus of biostatistics and computer science. Moreover, fostering reproducibility and replicability and promoting open science stands as a paramount objective of GBS 2.0.

In this manuscript, we described six areas of GBS that have emerged or matured since the original 2010 GBS paper. In addition to these areas, we mention two emerging areas likely to gain increasing prominence in the GBS of the future. The first is the use of artificial intelligence (AI) (e.g., OpenAI) both as an intervention and as a resource to enhance data processing and complex analysis of gerontology research data. While some AI technologies are currently deployed in healthcare settings (96), their role in quantitative research is still unknown. Another nascent application of modern computing power is in improving record linkage. While this technology has potential to enhance within-individual data integration for gerontology research, it can also increase risks of data de-identification and privacy violations unless privacy-preserving approaches are used (97), which may threaten the open science movement. While we await future versions of GBS that may harness these AI and record-linkage methods, we anticipate that GBS 2.0 will advance the training of a new cohort of clinical and epidemiological researchers specializing in research designs intended to improve healthspan and lifespan of older persons.

**Funding:** This work was supported by the National Institutes of Health (grant numbers R01 AG048069, R01 AG069915, R01 AG079854, RF1 NS128360, R03 AG070178, P30 AG0247, P30 AG021334, P30 AG066507, K25 AG068253, U54 AG063546, P30 AG021342, P30 AG066508, UL1 TR002014).

**Conflicts of Interest:** All authors are supported by grants or contracts from the National Institutes of Health.

**TABLE & FIGURE TITLES:**

**TABLE 1:** Challenges and Modern Statistical Methodologies for Gerontologic Research

**FIGURE 1:** Evolution from Gerontologic Biostatistics 1.0 to Gerontologic Biostatistics 2.0: A Visual Representation

**FIGURE 2:** Summary for Statistical Modeling and Machine Learning

# Table 1

| Statistical Challenge | Modern Methodology | Resources (with applications) | Utility |
|---|---|---|---|
| Advances in Experimental Design | Adaptive designs, platform trials, and dynamic regimes | SPIRIT Statement: (9)<br><br>NIH Collaboratory: (28, 29)<br><br>IMPACT Collaboratory: (30) | • Guidelines for clinical trial designs and protocols |
| Statistical Modeling and Machine Learning | Research question and data structure can guide choice of most appropriate analysis. | Overview: (25, 26)<br><br>Choice of approach: (27)<br><br>Gerontologic applications: (28, 30) | • Flexibility<br>• Optimality for Big Data<br>• Can use separately or together |
| Phenotypic Measurement Challenges | Signal intensive measurements and multidimensional geriatric outcomes | Cosine function bases (45)<br><br>Non-parametric functional regression and principal components analysis (46)<br><br>Joint analyses of multi-modal signals (47)<br><br>Joint and individual variance explained method (48)<br><br>The "index" rubric (52)<br><br>The "scale" rubric (53) | • Capture joint variation among the modes<br>• Capture variation specific to each mode<br>• Assess a multivariate measure |
| Analyzing -Omics Data: Microbiome | Compositional Data Analysis | Sparse Correlations for Compositional Data (74)<br><br>Robust Aitchison Distance (75)<br><br>Regularized or sparse centered or isometric log-ratio regression (72) | • Handle high-dimensional compositional data<br>• Enable adaptation of previously developed –omics workflows |
| | | Centered log ratio microbiome outcome models (73) | to microbiome data |
| Data Integration | Advanced methods for across-study/within study data integration and analysis | Generalized-meta analysis (COVID-19 mortality) (83, 84)<br><br>Integrate distributed data: Surrogate likelihood approach (88)<br><br>Empirical likelihood-based weighting for integrating secondary outcomes (90) | • Privacy-preserving<br>• Efficient communication<br>• Reliable statistical inference<br>Better prediction |
| Reproducibility, Replicability, and Open Science | Data and code sharing on accessible platforms | Best practices for data sharing (95)<br><br>Code sharing (GitHub) | • Facilitating rigor, reproducibility, and transparency<br>• Promoting open science |

**Figure 1**

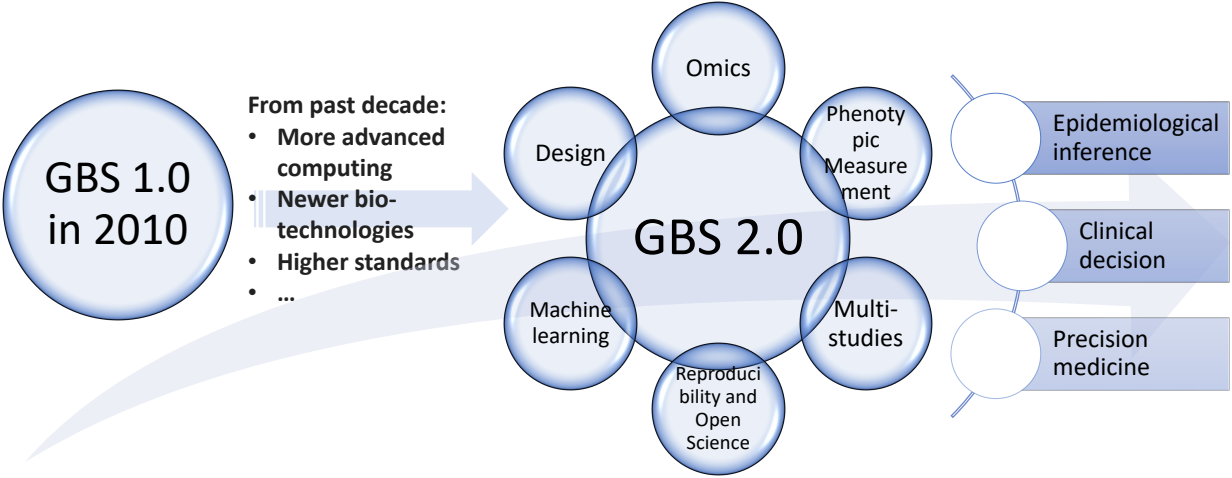

**Figure 2**

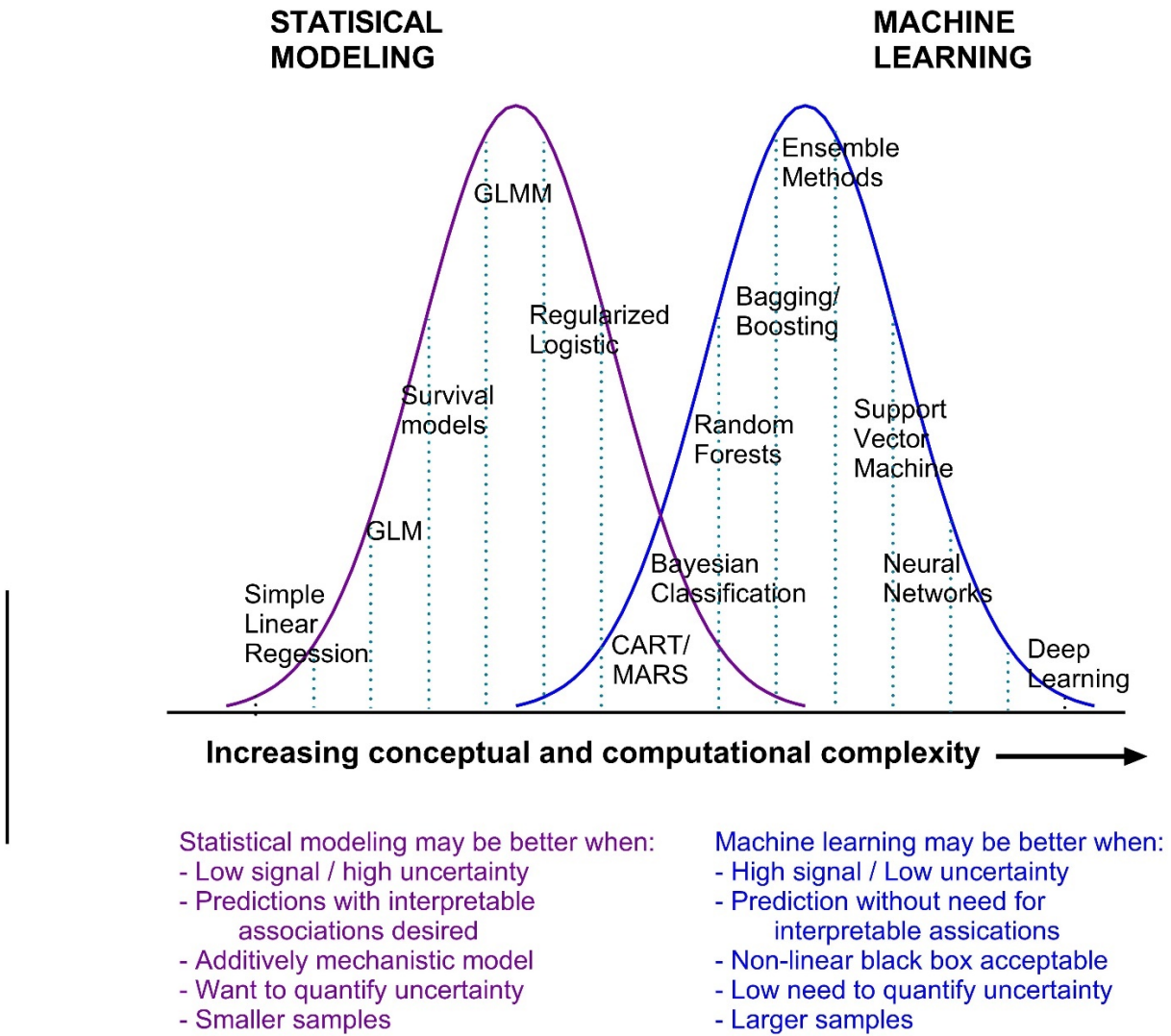